\documentclass[amsmath,amssymb,aps]{revtex4}
\usepackage{epsfig,dsfont,amssymb,amsmath,amsthm,amsfonts,amsbsy,mathrsfs} 
\usepackage{graphicx} 
\usepackage{color} 
\usepackage{bm}
\usepackage{multirow} 

\usepackage{natbib} 

\begin{document}

\title{Optimal escapes in active matter}
\author{Luca Angelani$^{1,2}$}
\email{luca.angelani@roma1.infn.it}
\affiliation{$^1$  Istituto dei Sistemi Complessi, Consiglio Nazionale delle Ricerche, Piazzale  A.  Moro  2,  I-00185  Roma,  Italy} 
\affiliation{$^2$ Dipartimento di Fisica, Sapienza Universit\`a di Roma, Piazzale  A.  Moro  2,  I-00185  Roma,  Italy}


\begin{abstract}
The out-of-equilibrium character of active particles, responsible for accumulation at boundaries in confining domains, determines not-trivial effects when considering escape processes. Non-monotonous behavior of exit times with respect to tumbling rate (inverse of mean persistent time) appears, as a consequence of the competing processes of exploring the bulk and accumulate at boundaries.
By using both 1D analytical results and 2D 
numerical simulations 
of run-and-tumble particles with different behaviours at boundaries,
we scrutinize this very general phenomenon of active matter, 
evidencing the role of accumulation at walls
for the existence of optimal tumbling rates for fast escapes.
\end{abstract}

\maketitle

\section{Introduction}
Consider a persistent random walk in a bounded domain. The motion is described by two parameters: the finite speed $v$ and the inverse of persistent time $\alpha$, also known as {\it tumbling rate} in run-and-tumble models 
\cite{Sch1993,weiss,maso1992,cates2012,mart,tai2008,Slo2016,Eva2018}.
Suppose there is a small aperture in the boundary allowing the walker to exit. We wonder about the existence of an optimal exit time with respect to system parameters. While a trivial decrease of exit times for faster walkers is evident, the role
of $\alpha$ is quite more difficult to evaluate. 
We expect that the peculiar  property of active matter 
to accumulate at boundaries 
plays a crucial role in determining the behavior of the system 
\cite{Bec2016,Ang2009,RDL2010,Sok2010,Gal2007,Rei2017,Wen2008,Elg2013,Elg2015,Cos2012,Cap2019,Moen2022,Mala2018,JPA2017,Bre2023}.
In general, we expect a competition between two characteristic times, the one spent by the walker in the bulk $\tau_{_{bulk}}$
, expected to grow with $\alpha$
(the more you tumble, the more you wander around), and the one spent on boundary $\tau_{_{boundary}}$, expected to decrease with $\alpha$ (tumbling promotes moving away from boundaries), 
giving rise to the possible existence of  
optimal tumbling rate values corresponding to minimum exit times.
We can give semi-quantitative arguments supporting this picture. We suppose that the time spent in the bulk by a persistent random walker 
before reaching the boundary is a growing function of $\alpha$ (assuming, on first approximation, linear dependence, in agreement with first-passage expressions \cite{Weiss1984,EPJE_FPT}),
\begin{equation}
    \tau_{_{bulk}} = a + b \ \alpha ,
\label{tau_bu}
\end{equation}
with $a$ and $b$ generic constants depending on $v$ and geometrical parameters. The time spent on boundary can be instead represented by a inverse proportionality:
\begin{equation}
   \tau_{_{boundary}} = \frac{c}{\alpha} ,
   \label{tau_bo}
\end{equation}
where we have assumed that the time spent on the boundary could diverge in the limit of null tumbling rate (infinite persistent time),  corresponding to a blocking situation on the walls (sticky boundaries).
Before exit, the particle spends part of the time  in the bulk and part in the boundary, 
so we can write the particle's lifetime as the sum of the two times,
$\tau = \tau_{_{bulk}} + \tau_{_{boundary}}$, getting the expression
\begin{equation}
  \tau = a + b \ \alpha + \frac{c}{\alpha} .
  \label{tauu}
\end{equation}
Therefore, we expect the existence of a minimum value 
$  \tau^* = a + 2 \sqrt{bc}$, 
obtained at a finite value of the tumbling rate  $\alpha^* = \sqrt{c/b}$.
Generalizing the argument to more complicated situations 
we can relax the hypothesis of complete blocking of particles at walls,
and write the exit time as
\begin{equation}
    \tau = \frac{f(\alpha)}{\alpha+\alpha_0} ,
    \label{tauu2}
\end{equation}
where $f(\alpha)$ 
is a smooth function of $\alpha$ \footnote{
By requiring that for  ($\alpha,v) \to \infty$ with finite 
$v^2/\alpha \sim  D$ 
(with $D$ the diffusion constant), one recovers the diffusive limit $\tau \propto D^{-1}$ \cite{Redner},
we can take $f$ as a polynomial quadratic function.
We can generalize (\ref{tauu2}) to generic power $\tau = f(\alpha) (\alpha +\alpha_0)^{-n}$,
with $f$ a polynomial function of order $n+1$.
}
and $\alpha_0$ is a parameter that depends on the properties of the boundary and the interactions between particle and wall.
We note that for $\alpha_0=0$ the exit time (\ref{tauu2}) can be put in the form (\ref{tauu}),
retrieving the perfect sticky boundary situation.
In the general case we  have that the optimal exit time is obtained at finite $\alpha^*$ satisfying 
$f(\alpha^*)=(\alpha^*+\alpha_0) f'(\alpha^*)$, but, in certain range of values of parameter $\alpha_0$,
it is reached at $\alpha^*=0$.
In other words, there is a critical value of the  parameter $\alpha_0$,
discriminating a region where the optimal escape corresponds to a finite tumbling rate from
a region where the fastest escapees are non-tumbling walkers.\\
These semi-quantitative arguments suggest that active random walks can exhibit very rich 
behaviors in escape processes, depending on the nature of the interaction with the boundary.
Recent experimental studies have shown  the importance of boundary interactions for the escape of
microalgae from circular pools \cite{Sou2022}.
Evidences of optimal escapes or optimal search strategies have been previously observed in numerical investigation of active particles in circular domains 
\cite{PAP2020,Deb2021,RBV2016,Zha2023}.
In this work we analyze in detail the role played by particle-boundary interactions for the occurrence
of optimal exit times of  run-and-tumble particles in bounded domains.
In particular, we will focus on one-dimensional and two-dimensional systems, studying different
boundary conditions. 
For the one-dimensional case we will exploit 
recent analytical results obtained for the run-and-tumble equations in the presence of 
partially absorption \cite{JPA2015,Bre2022}, sticky boundaries \cite{JPA2017,Bre2023}
and generic boundary conditions \cite{JPA2023}.
In Ref. \cite{JPA2023} a very general expression of the mean exit time was obtained, valid for a variety of different
types and combinations of boundaries, from reflecting to partially absorbing and sticky-like, resulting in non trivial
behaviors as a function of physical parameters, with the possible existence of non-monotonic trends 
in certain case studies.
Starting from these preliminary observations and results we 
conduct a detailed analysis and discuss in depth 
the conditions under which such a non-monotonic behaviors are present, 
evidencing the role of accumulation at boundaries by using tunable parameters to modulate its relevance.
The two-dimensional case will be studied considering circular domains with the presence of a 
narrow aperture on the boundary, that allows particles to escape.
By numerically investigating the particle dynamics for different particle-boundary interactions
(cases of complete, partial or absent alignment of the self-propelled orientation of the particle on the boundary) we will able to scrutinize the exit processes, elucidating the role of boundaries 
in determining optimal escapes.

\section{1D exact results}
\begin{figure}[t!]
\includegraphics[width=0.7\linewidth] {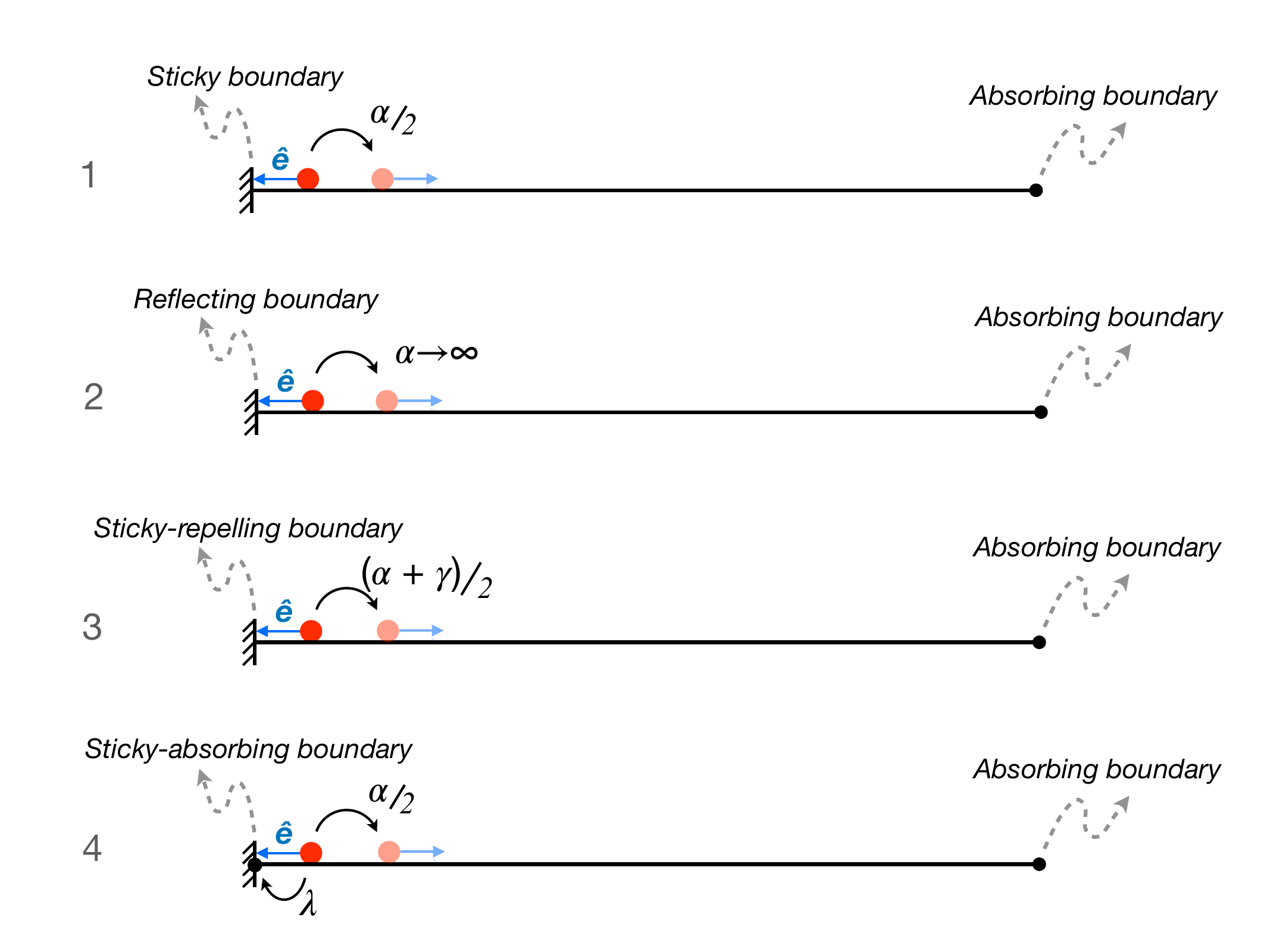}
\caption{\label{fig1}
Schematic representation of the analyzed cases corresponding to a 
1D run-and-tumble particle -- 
moving with velocity ${\mathbf v} = {\hat{\mathbf{e}}} v$ ($\hat{\mathbf e} = \pm {\hat{\mathbf x}}$)
and tumbling rate $\alpha$ --
in a finite domain with an absorbing boundary on the right side 
of the interval and 
different boundary conditions on the left side.
1) Sticky boundary: the particle remains stuck at the boundary until 
it reverses its direction of motion with rate $\alpha/2$
(the factor $1/2$ comes from the fact that, after a tumble, 
the particle can, with equal probability, maintain its direction of motion or reverse it). 
2) Reflecting boundary: when the particle hits the wall it instantaneously reverses its direction of motion (corresponding to an infinite tumbling rate at the wall).
3) Sticky-reflecting boundary: the particle remains stuck at the boundary until 
it reverses its direction of motion with rate $(\alpha+\gamma)/2$.
4) Sticky-absorbing boundary: the particle at the boundary can reverse its direction of motion (with rate $\alpha/2$) or be absorbed (with rate $\lambda$).}
\end{figure}

\begin{figure}[t!]
\includegraphics[width=0.7\linewidth] {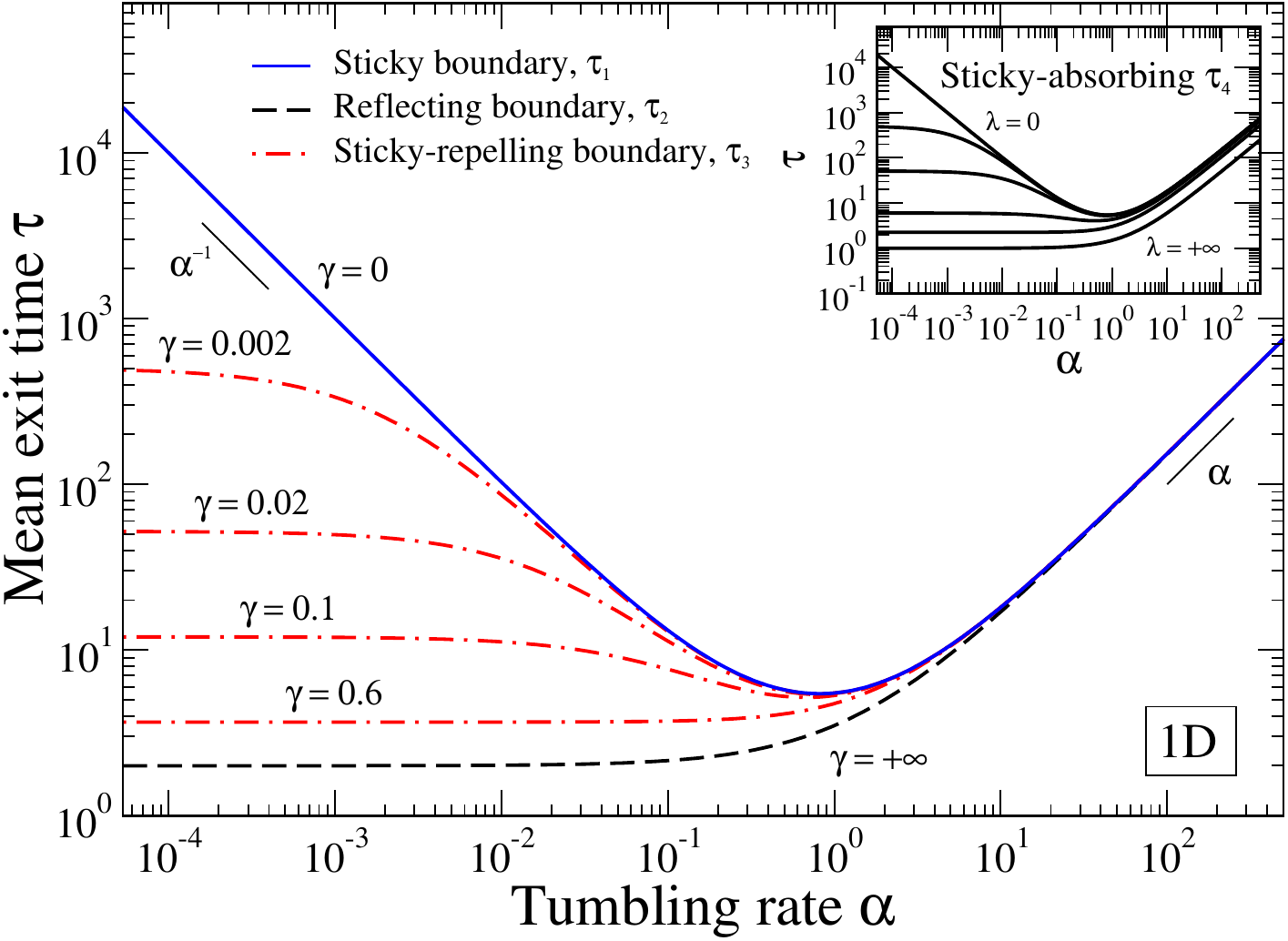}
\caption{\label{fig2}
Mean exit times $\tau$ as a function of the tumbling rate $\alpha$ for a 1D run-and-tumble particle in a finite interval $[-R,R]$ with 
absorbing boundary at $x=R$. Different boundary conditions are considered at $x=-R$.
$\tau_1$ corresponds to sticky boundary (see expression (\ref{tau1}) in the text), $\tau_2$ to reflecting boundary 
(\ref{tau2}), $\tau_3$ to sticky-repelling boundary with enhanced tumbling rate at wall $\alpha+\gamma$ (\ref{tau3}).
Inset. Exit times for the case of sticky-permeable wall at $x=-R$, with exit rate $\lambda$ (\ref{tau4});
the reported cases correspond to (from up to bottom) $\lambda=0, 10^{-3},10^{-2},10^{-1},0.4,+\infty$.
We set $v=R=1$ ($\gamma_c \simeq 0.549$, $\lambda_c \simeq 0.366$).
}
\end{figure}

\noindent
We consider a run-and-tumble particle, with speed $v$ and tumbling rate $\alpha$, confined in a 1D segment $(-R,R)$ 
\cite{Sch1993,weiss,maso1992,cates2012,mart,tai2008,Slo2016,Eva2018,JPA2017,JPA2015}.
We assume that the particle starts its motion at the origin $x=0$.
Boundary conditions are as follows. There is an hard wall at $x=-R$, allowing particle accumulation, i.e., the particle gets stuck to the wall until a tumble event reverses its direction of motion \cite{JPA2017,Bre2023}.
An absorbing barrier is present at the exit point $x=R$. 
A schematic representation of the system is shown in Fig.\ref{fig1} (case $1$).
Following similar analysis of Ref. \cite{JPA2017},
where only symmetric cases were taken into account, 
it is possible to obtain the exact expression of the mean exit time, i.e. the mean first passage time for the particle to reach the exit point $x=R$ -- 
see the Appendix and \cite{JPA2023}, 
in which a general treatment is given for generic boundaries --
obtaining
(Fig.\ref{fig2}, blue full line)
\begin{equation}
\label{tau1}
\tau_1 = 3\ \frac{R}{v} + \frac{3}{2} \frac{R^2}{v^2} \ \alpha + \frac{1}{\alpha}  .
\end{equation}
This expression is exactly what expected by qualitative arguments considering the exit time as a
sum of bulk-time (\ref{tau_bu}) and 
boundary-time (\ref{tau_bo}), with $a=3R/v$, $b=3R^2/2v^2$ and $c=1$.
The mean exit time diverges as $\alpha^{-1}$ and $\alpha$ 
at small and high tumbling rates, respectively (see Fig.\ref{fig2}).
Therefore, there exists a minimum value 
$~{\tau_1^*=(3+\sqrt{6}) R/v}$
obtained for the optimal tumbling rate 
$~{\alpha^*=\sqrt{2/3}\ v/R}$.

In order to study the role of particle-boundary interaction in determining 
the presence of an optimal exit time at finite $\alpha$
we now consider different kinds of boundaries. Let us first assume 
a totally reflecting boundary at $x=-R$. In this case, arriving at the boundary, the particle, is no more stuck at wall, but it reverses instantaneously its direction of motion
(Fig.\ref{fig1}, case 2).
Now there is no more accumulation at boundaries and the exit time has the following form 
-- see the Appendix and Refs. \cite{JPA2015,JPA2023} --
\begin{equation}
\tau_2 = 2 \frac{R}{v}
+ \frac32 \frac{R^2}{v^2}\ \alpha .
\label{tau2}
\end{equation}
In other words the boundary-time vanishes ($c=0$),
the exit time becomes a simple growing function of $\alpha$
and its minimum value $\tau^*_2=2R/v$ is reached at $\alpha^*=0$
(see Fig.\ref{fig2}, black dashed curve),
with only the presence of a crossover when the run length $v/\alpha$ is comparable 
with the system size $R$.

We now analyze an intermediate situation between the previous two.
We generalize the first analyzed case, by considering that a particle stuck at the $x=-R$ boundary reorients its self-propelling direction at a rate $\alpha+\gamma$ instead of $\alpha$,
i.e., the tumbling rate at the wall is higher than that on the bulk, reducing the stickiness of the wall (Fig.\ref{fig1}, case 3).
For $\gamma=0$ we recover the first case, while for $\gamma \to \infty$ we get the reflecting case.
The parameter $\gamma$, as we will see,  allows us to switch between different behaviors, suppressing the presence of optimal exit times at finite $\alpha$ for certain parameter values.
In this case we can show that the mean exit time is 
-- see the Appendix and Ref.s \cite{JPA2017,JPA2023} --
\begin{equation}
    \tau_3 = 2 \frac{R}{v}
+ \frac32 \frac{R^2}{v^2}\ \alpha +
\frac{1+\alpha R/v}{\alpha+ \gamma} .
\label{tau3}
\end{equation}
It is evident here a more complex form of the exit time,
due to the non trivial particle-boundary interaction at wall.
The above expression interpolates the previous two: for $\gamma=0$ we get (\ref{tau1}), while 
for $\gamma \to \infty$ we recover (\ref{tau2}).
We also note that (\ref{tau3}) can be cast in the form (\ref{tauu2}), with $\alpha_0=\gamma$.
Analyzing the behavior of $\tau_3$ as a function of the tumbling rate, 
we find that the condition for the existence of an optimal exit time at finite $\alpha^*$ is 
\begin{equation}
    \gamma < \gamma_c =\frac{v}{R} \frac{\sqrt{7}-1}{3} ,
\end{equation}
and we obtain
$~{\alpha^*=-\gamma +\sqrt{2/3} \sqrt{(1-R\gamma/v)}\ v/R}$ and
$~{\tau_3^* = [1+(\alpha^*+\gamma/2)R/v]\ 3R/v}$.
For  $\gamma> \gamma_c$, instead,  the minimum exit time 
$\tau_3^* = 2R/v+1/\gamma$
is reached by non-tumbling particles ($\alpha^*=0$),
due to the less relevance of the time spent at wall.
In Fig.\ref{fig2} we report some examples of exit times for different values of the parameter $\gamma$, above and below the critical value $\gamma_c$ (red dot-dashed curves).

The last non trivial situation we consider is that of a partially permeable wall at $x=-R$, allowing particles to exit with rate $\lambda$. Now the particle can exit, as before, reaching the exit point $x=R$, but also with rate $\lambda$ when it is stuck at $x=-R$ (Fig. \ref{fig1}, case 4),
then reducing the sticky property of the wall.
In this case the exit time reads -- see the Appendix and Ref.s \cite{JPA2017,JPA2023} --
\begin{equation}
    \tau_4 = \frac{R}{v}
+ \frac12 \frac{R^2}{v^2}\ \alpha +
\frac{(1+\alpha R/v)^2}{\alpha+ 2 \lambda (1+\alpha R/v)} .
\label{tau4}
\end{equation}
We note that for $\lambda=0$ we retrieve the  expression of impermeable wall (\ref{tau1}),
while, for $\lambda \to \infty$, we obtain the first-passage time in the presence of 
two absorbing boundaries, $\tau_4 = R/v + R^2\alpha/2v^2$ 
\cite{EPJE_FPT,JPA2015,JPA2023}.
Also in this case the expression (\ref{tau4}) can be cast in the form (\ref{tauu2}), 
with $\alpha_0=(1/2\lambda + R/v)^{-1}$.
We have that the condition for the existence of an optimal exit time 
$\tau^*$ at finite $\alpha^*$, is
\begin{equation}
    \lambda < \lambda_c =\frac{v}{R} \frac{\sqrt{3}-1}{2} .
\end{equation}
For  $\lambda> \lambda_c$ the minimum exit time is at $\alpha^*=0$.
In Fig.\ref{fig2} (inset) we report the exit times (\ref{tau4}) for different values of the parameter $\lambda$.
\\
Summarizing, we can be conclude that the accumulation at the wall is responsible for the increase in the time
that particles spend on the boundary as the tumbling rate decreases, in contrast to the opposite trend of the time spent by particles in the bulk.
This results in the existence
of optimal escape times at a finite value of the tumbling rate $\alpha^*$.
When boundary accumulation is progressively inhibited, for example increasing the 
tumbling rate on the boundary or allowing particle absorption at wall, 
the optimal $\alpha^*$ begins to decrease and finally reaches zero
at a certain value of the additional parameters 
($\gamma$ or $\lambda$ in the previous models)
that describe the inefficiency of the wall in allowing particles accumulation.

\section{2D numerical results}
We now turn to analyze the case of planar motions.
We consider a 2D run-and-tumble particle in a circular
domain of radius $R$. 
The particle moves at constant speed $v$ (in the bulk) and reorients its direction
of motion at rate $\alpha$, with the tumbling angle chosen from a uniform distribution.

\begin{figure}[t!]
\includegraphics[width=0.7\linewidth] {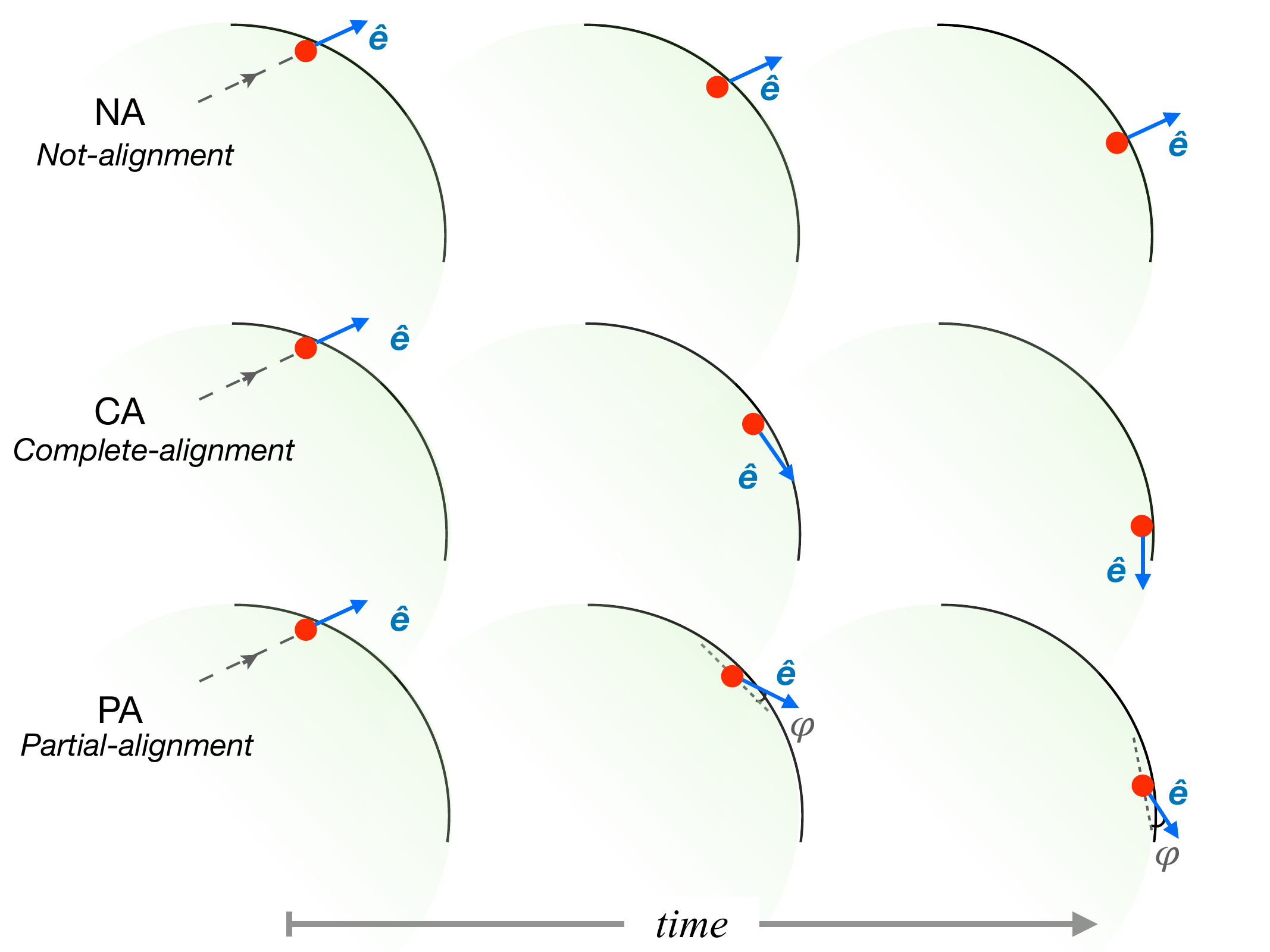}
\caption{\label{fig3}
Schematic representation of the different  boundary-particle interactions 
considered for a 2D run-and-tumble particle inside a circular domain. The upper case corresponds to the 
not-alignment situation (NA), in which the particle, hitting the edge, does not change its self-propelling orientation 
${\hat{\mathbf e}}$ and continues to move (with decreasing speed) along the border until its orientation is parallel to the boundary normal (unless it has tumbled or encountered the exit in the mean- time). The second case corresponds to an (instantaneous) complete-alignment (CA) of the self-propelling orientation with the boundary wall. The particle continues its motion along the boundary at speed $v$. The last case refers to the partial-alignment (PA) of  ${\hat{\mathbf e}}$ along the boundary, with a finite value of the angle 
$\varphi$ between it and the tangent to the boundary. The particle proceeds at speed 
$v \cos \varphi$.
}
\end{figure}
A narrow aperture on the boundary allows the particle to exit. 
We will examine three case studies, corresponding to different particle-boundary interactions. 
The first case is that of a particle that does not change its orientation when arriving at the border. This is the case, for example, of spherical active particles, 
where torques are absent \cite{How2007,Bec2016}.
We refer to this case as not-alignment case (NA).
The second case is that of a complete alignment (CA) of particle orientation along the boundary. When colliding to the border the particle changes instantaneously its self-propelling direction of motion parallel to the wall. This is the case, for example, of elongated particles, such as {\it E.coli} bacteria  \cite{Ecoli}
(we are neglecting the transient time for the complete alignment 
of the particle orientation to the boundary).
The last case  analyzed is that of partial alignment (PA), corresponding to a particle that maintains a fixed angle $\varphi$ between its orientation  and the tangent vector to the boundary. This happens, for example, in the case of sperm cells, where the extension of flagellum prevent a complete alignment of the cell along the boundary \cite{Elg2010,Den2012}.
Summarizing, we are considering a particle that, when arriving at the boundary, proceeds its motion along it with: decreasing speed in the NA case, with speed $v$ in the CA case, and with speed $v \cos{\varphi}$ in the PA case. 
The particle motion ends when it encounters the aperture and exits from the domain 
(we consider point-like particles that exit as soon as they cross the exit zone, 
if they come from the bulk,
or touch its edge, if they come from the circular boundary).
A schematic representation of the analyzed cases is shown in Fig.\ref{fig3}.
We use numerical simulation to investigate such active random walks in 2D circular geometry, proceeding as follow. We start with a particle at the center of the circular domain. 
We then sequentially extract directions of motion with uniform angular distribution and rum times from
an exponential distribution $\alpha \exp{(-\alpha t)}$.
The particle moves at constant speed $v$ along straight lines in the bulk and, when it hits the boundary, it proceeds along the border for the rest of the run time,
proceeding with a speed which depends on the case analyzed (as described before).
The motion of the particle  ends when it reaches  the small aperture and exits the domain.
Average over $10^4$ up to $10^5$ runs are considered.\\
\begin{figure}[t!]
\includegraphics[width=0.7\linewidth] {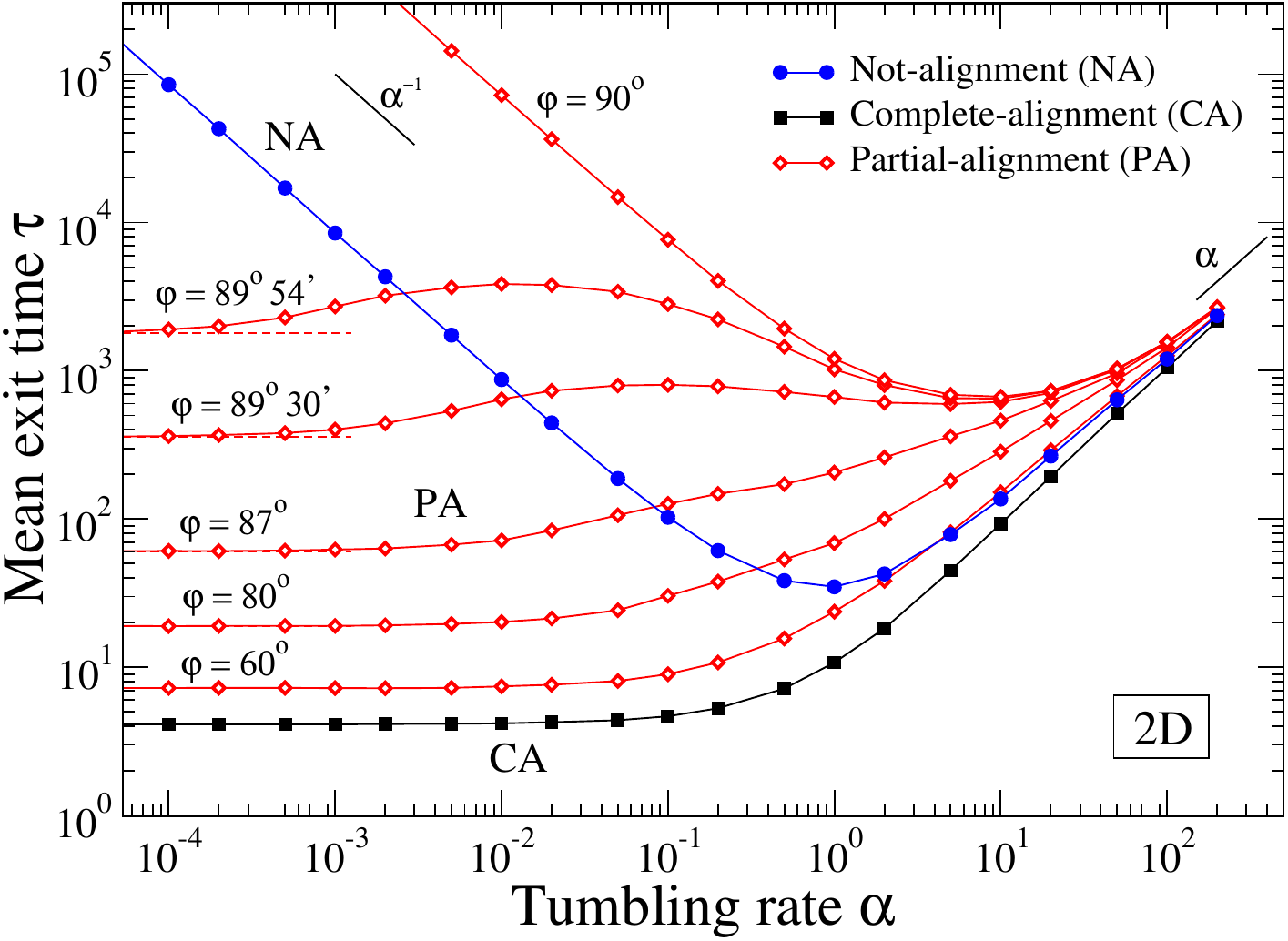}
\caption{\label{fig4}
Mean exit times $\tau$ as a function of the tumbling rate $\alpha$ for a 2D run-and-tumble particle
inside a disk of radius $R$ with a small aperture on the boundary of size $\delta$, with $\delta/R=\pi/180$ (angular aperture of $1^{\text{o}}$).
The reported cases correspond to not-alignment of particle velocity along the boundary (NA, blue full circles),
complete alignment (CA, black full squares), finite angle $\varphi$ between the particle orietation and 
the tangent to the boundary (PA, red open diamonds) 
-- see  the schematic pictures in Fig.\ref{fig3}. 
The lines are guides to the eye.
The asymptotic limit $\alpha \to 0$ of PA curves is
$\tau(0) = R/v [1+(2\pi-\delta/R)^2/(4\pi \cos\varphi)]$ (dashed lines, see Supplemental Material).
We set $v=R=1$.}
\end{figure}
In Fig.\ref{fig4} we show the mean exit times as a function of the tumbling rate $\alpha$ for the different cases analyzed.
It is evident the presence of a minimum at finite values of $\alpha$ in the NA case and PA case for sufficiently high values of
the alignment angles $\varphi$ (close to $\pi/2$). 
We note that the PA case with $\varphi=\pi/2$ shows a divergent exit time at small
$\alpha$, similarly to the NA case,
because of the perfect sticky condition that occurs at the boundary in these cases.
For smaller angles, down to the case of complete alignment (PA with $\varphi = 0$, corresponding to CA),
the minimum exit time is reached by non-tumbling particles ($\alpha=0$). 
The reported trends are very similar to the cases analyzed 
in 1D, highlighting the ability of the simplified one-dimensional models to effectively capture 
the essential aspects of the physics of the problem. However, we note that the specific functional dependencies in 1D and 2D can be different, and, for example,
the $\tau(\alpha)$ trends (see the PA curves in the figure) cannot always be captured by a simplified expression like Eq.(\ref{tauu2}), requiring, instead, more involved functional forms
(see the footnote at page 1).
Summarizing, we can then say that, even in the two-dimensional case, the accumulation  
at boundaries determines the existence of optimal exit times at finite values of the tumbling rate,
which eventually becomes zero when sticky properties of boundaries are reduced.

\section{Conclusions}
We have investigated the escape processes of confined active particles. 
In particular, we focused on the possible existence of optimal 
escape times. We found that the peculiar property of active matter to accumulate at walls gives rise to the existence of optimal tumbling rates corresponding to fast escapes.
By introducing four different kinds of boundary conditions
in 1D run-and-tumble models, we were able to obtain analytical expressions of escape times, which allow us to discuss the relevance of accumulation processes at boundaries for the existence
of optimal finite values of tumbling rates.
These findings are corroborated by the analysis of 2D run-and-tumble particles confined 
in a circular domain. By numerically studying the escape dynamics for different kinds of
particle-boundary interactions (not/complete/partial - alignment of particle's self-propulsion with
the boundary wall) we find again the existence of optimal tumbling rate values for fast escapes,
which tend to zero (the particles that come out the fastest are the ones that do not tumbling)
by inhibiting the sticky properties of the walls.
We expect this is a very general behavior of active matter, 
occurring whenever particles accumulation is present.
Optimal escapes are, in fact,  present also in physical situations where the escape (absorbing) regions are in the bulk and not along the boundary \cite{RBV2016}, 
or in the case in which the confinement of particles is due to external fields 
(potential barriers) instead of geometrical  constraints 
\cite{Cap2021,Mil2021}.
Indeed, also in these cases,
we have particles accumulation and the presence of competition between  times spent on the bulk and on the boundary, which we have demonstrated are essential ingredients to have  optimal  escapes.
Possible directions for future investigations could be to explore different geometries 
of the confining box, with the possible presence of curvature-dependent accumulation
\cite{Fily2014}, to analyze the  differences with respect to other active particles models,
such as the active Brownian particle model \cite{Sol2015}, 
or to extend the investigation to three-dimensional domains
and including particle-particle interactions.
Finally, it would be interesting to investigate optimal escapes in experiments, for example, 
by devising  genetically modified bacteria, 
with tunable tumbling rate controlled by external fields,
in analogy to light-controlled speed in photokinetic bacteria \cite{photok}.
The challenge here, in addition to synthesize these new kind of bacteria,
is to develop an experimental set-up to study microswimmers 
in confined environments with the presence of narrow  apertures enabling escape,
with all the difficulties associated with the presence of effects not 
easily controlled at these microscales, such as hydrodynamic interactions with  bourdaries.
Alternatively, one could use non-living particles, such as  shaped active Brownian colloids  with controlled rotational dynamics \cite{FR2020}
or the recently employed commercially toy robots Hexbugs \cite{Volpe2022}.

\acknowledgments
I acknowledge financial support from the Italian Ministry of University and Research (MUR) under 
PRIN2020 Grant No. 2020PFCXPE.

\section{Appendix}
We derive here the expressions of the mean exit times of a run-and-tumble particle in a bounded 
one-dimensional domain.
The probability density function $P(x,t)$ to find the particle at position $x$ at time $t$ 
obeys the so-called telegrapher's equation \cite{weiss}
\begin{equation}
  ( v^2 \partial_x^2 - \partial_t^2 - \alpha \partial_t ) P = 0 ,
\end{equation}
where $v$ is the particle's speed, $\alpha$ its tumbling rate (inverse of the persistence time) and we denote with 
$\partial_t=\partial/\partial t$ and
$\partial_t^2=\partial^2/\partial t^2$ the first- and second-derivative operators.
Assume that the particle starts its motion at the origin $x=0$ and that the domain extends from $-R$ to $R$.
The Laplace transformed function ${\tilde P}(x,s)=\int_0^\infty dt \exp(-st) P(x,t)$ obeys the equation
\begin{equation}
    [v^2 \partial_x^2 -s(s+\alpha) ] {\tilde P} = - (s+\alpha) \delta(x)  .
\label{tel_eq_L}
\end{equation}
We assume that there is an absorbing boundary at $x=R$, corresponding to the boundary condition 
-- see equation (24) of \cite{JPA2015} with $\epsilon=1$ --
\begin{equation}
   ( v \partial_x + s+\alpha ) {\tilde P} |_{x=R} = 0 .
\end{equation}
In this work we consider different types of boundary at $x=-R$, see figure \ref{fig1}.
The first one is a sticky boundary, where the particle, once arriving on it, remains stuck until
a tumble event reverses its direction of motion. In this case the boundary condition reads 
 -- see equation (11) in \cite{JPA2017} --
\begin{equation}
   ( v \partial_x - s ) {\tilde P} |_{x=-R} = 0 .
\label{bc1}
\end{equation}
The second case  we consider is that of a reflecting boundary, where the particle reverses instantaneously its direction of motion. Now the boundary condition becomes -- see equation (23) of \cite{JPA2015} with $\epsilon=0$ --
\begin{equation}
   v \partial_x {\tilde P} |_{x=-R} = 0 .
\label{bc2}
\end{equation}
The third case is that of a sticky-repelling boundary, where the stuck particle reorients its direction of motion with a rate $\alpha+\gamma$ (with $\gamma>0$), higher than that of the particle in the bulk $\alpha$.
The boundary condition now reads 
-- see equation (25) of \cite{JPA2017} with $\alpha_{_W}=\alpha+\gamma$ --
\begin{equation}
   [ v \partial_x - s (s+\alpha)/(s+\alpha+\gamma) ] {\tilde P} |_{x=-R} = 0 .
\label{bc3}
\end{equation}
The last case considered is a sticky-absorbing boundary, where the stuck particle can be
absorbed with rate $\lambda$. The corresponding boundary condition is 
-- see equation (33) in \cite{JPA2017} -- 
\begin{equation}
   [ v \partial_x - (s+\alpha) (s+\lambda)/(s+\alpha+\lambda) ] {\tilde P} |_{x=-R} = 0 .
\label{bc4}
\end{equation}
The solutions of eq. (\ref{tel_eq_L}) are given by the superposition of  exponential functions 
$\exp(\pm c x)$, with $v^2 c^2 = s(s+\alpha)$, and with coefficients  obtained by imposing
boundary conditions and the 
continuity (discontinuity) of ${\tilde P}$ ($\partial_x {\tilde P}$) at $x=0$ \cite{JPA2015,JPA2017}.
Once obtained  the solutions ${\tilde P}$ for the different boundary conditions analyzed,
the probability distribution $\varphi(t)$ of the exit time is obtained as the (minus) time-derivative of the survival probability ${\mathbb P} = \int_{-R}^{+R} dx \ P + W$, where $W$ is the probability 
to find the particle stuck at $x=-R$ in the case of sticky-like property of the boundary
\cite{JPA2017,JPA2023}.
There is a relation between $W$ and $P|_{x=-R}$. For example, in the case of sticky boundary,
from the continuity equation $\partial_t W = - J|_{x=-R}$ ($J$ is the current $v P_+ - v P_-$, with 
$P_{\pm}$ the right- and left-oriented particle distribution) and the boundary relation 
$vP_+|_{x=-R} = \alpha W/2$ (the flow of right moving particles  at the left boundary is given 
by the fraction of stuck particles that reverse their direction of motion) \cite{JPA2017},
we obtain  $(\partial_t +\alpha)W=vP|_{x=-R}$, or, in the Laplace domain 
$(s+\alpha){\tilde W}=v{\tilde P}|_{x=-R}$.
In the case of reflecting boundary there are no stuck particles, i.e., $W=0$.
For sticky-repelling boundary we have that 
$\partial_t W = - J|_{x=-R}$ and $vP_+|_{x=-R} = (\alpha +\gamma)W/2$ 
\cite{JPA2017}, leading to the relation
$(s+\alpha+\gamma){\tilde W}=v{\tilde P}|_{x=-R}$.
For the fourth case analyzed, the sticky-absorbing boundary, from 
$\partial_t W = - J|_{x=-R} - \lambda W$ and $vP_+|_{x=-R} = \alpha W/2$ 
\cite{JPA2017}, we obtain 
$(s+\alpha+\lambda){\tilde W}=v{\tilde P}|_{x=-R}$.
The expression of the mean exit time distribution can be obtained
by using ${\tilde {\varphi}} = 1 - s {\tilde {\mathbb P}} $, leading to
${\tilde {\varphi}} = v {\tilde P}|_{x=R} + v h {\tilde P}|_{x=-R}$,
where $h=0$ for the first three boundaries  and  $h=\lambda/(s+\alpha+\lambda)$ 
for the fourth case.
The mean exit time can be finally obtained
as $\tau = \int_0^\infty  t \varphi dt =  -\partial_s {\tilde{\varphi}}|_{s=0}$. \\
For the sake of simplicity we report here only the explicit solution for the first kind of boundary,
i.e., the sticky boundary (\ref{bc1}). For the other types of boundaries one can repeat exactly the same procedure to obtain the final expressions of the mean exit times given in the main text.
For boundary (\ref{bc1}) we have that the solution of (\ref{tel_eq_L}) reads
${\tilde P} = A_1 e^{cx} + A_2 e^{-cx}$  for $x>0$ and 
${\tilde P} = B_1 e^{cx} + B_2 e^{-cx}$  for $x<0$, where the coefficients are
\begin{eqnarray}
    A_1 &=& (c/2s) (vc-s-\alpha) e^{-cR}  F/ Q \nonumber ,\\ 
    A_2 &=& (c/2s) (vc+s+\alpha) e^{+cR} F/ Q \nonumber  ,\\ 
    B_1 &=& (c/2s) (vc+s) e^{+cR} G / Q  \nonumber ,\\ 
    B_2 &=& (c/2s) (vc-s) e^{-cR} G / Q  \nonumber , 
\end{eqnarray}
and 
\begin{eqnarray}
F &=& vc \cosh(cR) + s \sinh(cR) \nonumber ,\\
G&=& vc \cosh(cR) + (s+\alpha) \sinh(cR) \nonumber  ,\\
Q &=& vc (2s+\alpha) \cosh(2cR) \nonumber \\
&+& [v^2 c^2 + s(s+\alpha)] \sinh(2cR) \nonumber  .
\end{eqnarray}
The exit time distribution is in this case
\begin{equation}
    {\tilde{\varphi}} = (s+\alpha) F / Q ,
\end{equation}
and the mean exit time reads
\begin{equation}
    \tau = \frac{3R}{v}  + \frac{3R^2\alpha}{2v^2} + \frac1\alpha  ,
\end{equation}
that is the expression (\ref{tau1}) reported in the main text.
Similar derivations can be made for the other boundaries considered in this work,
obtaining the corresponding expressions of the mean exit times given in the main text.
A very general and complete derivation, considering generic boundary conditions, is given in
\cite{JPA2023}.

%

\begin{thebibliography}{99}


\bibitem{Sch1993}
Schnitzer M. J.
{\it Phys. Rev. E}{48}{1993}{2553}.

\bibitem{weiss}
Weiss G.H.
{\it Phys. A (Amsterdam, Neth.)} {311} {2002} {381}.

\bibitem{maso1992}
Masoliver J., Porr\`a J. M. \and Weiss G. H.
{\it Phys. Rev. A} {45} {1992} {2222}.

\bibitem{cates2012}
Cates M. E.
{\it Rep. Prog. Phys.} {75} {2012} {42601}.

\bibitem{mart} 
Martens K., Angelani L., Di Leonardo R. \and Bocquet L.
{\it Eur. Phys. J. E} {35} {2012} {84}.

\bibitem{tai2008} 
Tailleur J. \and Cates M. E.
{\it Phys. Rev. Lett.}  {100} {2008} {218103}.

\bibitem{Slo2016}
Slowman A. B., Evans M. R. \and Blythe R. A.
{\it Phys. Rev. Lett.} {116} {2016} {218101}.

\bibitem{Eva2018}
Evans M. R. \and Majumdar S. N.
{\it J. Phys. A: Math. Theor.} {51} {2018}  {475003}.


\bibitem{Bec2016}
Bechinger C. {\it et al.}
{\it Rev. Mod. Phys.} {88} {2016} {045006}.


\bibitem{Ang2009}
Angelani L., Di Leonardo R. \and  Ruocco G.
{\it Phys. Rev. Lett.} {102} {2009} {048104}.

\bibitem{RDL2010}
Di Leonardo {\it et al.}
{\it Proc. Natl. Acad. Sci.} {107} {2010} {9541}.

\bibitem{Sok2010}
Sokolov A.  {\it et al.}
{\it Proc. Natl. Acad. Sci.} {107} {2010} {969}.

\bibitem{Gal2007}
Galajada P., Keymer J., Chaikin P. \and Austin R.
{\it J. Bacteriol.} {189} {2007} {8704}.

\bibitem{Rei2017}
Reichhardt  C. J. O. \and Reichhardt  C.
{\it Annu. Rev. Condens. Matter Phys.} {8} {2007} {51}.

\bibitem{Wen2008}
Wensink H. H. \and L\"owen H.
{\it Phys. Rev. E} {78} {2008} {031409}.

\bibitem{Cos2012}
Costanzo  A., Di Leonardo  R., Ruocco  G. \and Angelani  L.
{\it J. Phys.: Condens. Matter} {24} {2012} {065101}.


\bibitem{Elg2013}
Elgeti  J. \and Gompper G.
{\it Europhys. Lett.} {101} {2013} {48003}.

\bibitem{Elg2015}
Elgeti  J. \and Gompper G.
{\it Europhys. Lett.} {109} {2015} {58003}.


\bibitem{Cap2019}
Caprini  L. \and Marconi  U. M. B.
{\it Soft Matter} {15} {2019} {2627}.


\bibitem{Moen2022}
Moen  E. Q. Z., Olsen  K. S., R{\o}nnig  J. \and Angheluta  L.
{\it Phys. Rev. Research} {4} {2022} {043012}.

\bibitem{Mala2018}
Malakar  K. {\it et al.}
{\it J. Stat. Mech.} {} {2018} {043215}.

\bibitem{JPA2017}
Angelani L.
{\it J. Phys. A: Math. Theor.} {50} {2017} {325601}.

\bibitem{Bre2023}
Bressloff P. C.
{\it J. Stat. Mech.} {} {043208} {2023} 




\bibitem{Weiss1984}
Weiss G. H.
{\it J. Stat. Phys.} {37} {1984} {325}.

\bibitem{EPJE_FPT}
Angelani  L., Di Leonardo  R. \and Paoluzzi  M.
{\it Eur. Phys. J. E} {37} {2014} {59}.

\bibitem{Redner}
Redner S 2001
{\it A Guide to First-Passage Processes} (Cambridge University Press, Cambridge, UK)





\bibitem{Sou2022}
Souzy M. {\it et al.}
{\it Phys. Rev. Research} {4} {2022} {L022029}.


\bibitem{PAP2020}
Paoluzzi  M., Angelani  L. \and Puglisi  A.
{\it Phys. Rev. E} {102} {2020} {042617}.

\bibitem{Deb2021}
Debnath T. {\it et al.}
{\it J. Chem. Phys.} {155} {2021} {194102}.

\bibitem{RBV2016}
Rupprecht  J.F., B\`enichou  O. \and Voiturez  R.
{\it Phys. Rev. E} {94} {2016} {012117}.

\bibitem{Zha2023}
Zhang W., Li Y., Marchesoni F., Misko V. R. \and Ghosh  P. K.
{\it Entropy} {25} {2023} {271}.

\bibitem{Bre2022}
Bressloff P. C.
{\it J. Stat. Mech.} {} {113206} {2022} 


\bibitem{JPA2015}
Angelani  L.
{\it J. Phys. A: Math. Theor.} {48} {2015} {495003}.



\bibitem{JPA2023}
Angelani L.
{\it J. Phys. A: Math. Theor.} {56} {2023} {455003} 


\bibitem{How2007}
Howse J. R.  {\it et al.}
{\it Phys. Rev. Lett.} {99} {2007} {048102}.


\bibitem{Ecoli}
Berg H C 2004
{\it E.coli in Motion} (Springer-Verlag, New York).


\bibitem{Elg2010}
Elgeti  J., Kaupp  U. B. \and Gompper  G.
{\it Biophys. J} {99} {2010} {1018}.

\bibitem{Den2012}
Denissenko  P., Kantsler  V., Smith  D. J. \and Kirkman-Brown J.
{\it Proc. Natl. Acad. Sci. U.S.A.} {109} {2012} {8007}.




\bibitem{Cap2021}
Caprini  L., Cecconi  F. \and Marconi  U. M. B.
{\it J. Chem. Phys.} {155} {2021} {234902}.


\bibitem{Mil2021}
Militaru A. {\it et al.}
{\it Nat. Commun.} {12} {2021} {2446}.

\bibitem{Fily2014}
Fily Y. , Baskaran A. \and Hagan M.F.
{\it Soft Matter} {10} {2014} {5609}.

\bibitem{Sol2015}
Solon A.P., Cates M.E. \and Taileeur J.
{\it Eur. Phys. J. Special Topics} {224} {2015} {1231}.


\bibitem{photok}
Frangipane  G. {\it et al.}
{\it ELife} {7} {2018} {e36608}.

\bibitem{FR2020}
Fernandez-Rodriguez M. A. {\it et al.}
{\it Nat. Commun.} {11} {2020} {4223}.

\bibitem{Volpe2022}
Balda A.B., Argun A., Callegari A. \and Volpe G.
{\it arXiv:2209.04168}{}  {2022} {}.


\end{thebibliography}
\end{document}